\documentclass[12pt]{article}
\usepackage{epsf}
\def\be{\begin{equation}}
\def\ee{\end{equation}}

\def\ltap{\raisebox{-.4ex}{\rlap{$\sim$}} \raisebox{.4ex}{$<$}}

\begin{document}
\begin{titlepage}
\begin{center}
{\Large \bf William I. Fine Theoretical Physics Institute \\
University of Minnesota \\}
\end{center}
\vspace{0.2in}
\begin{flushright}
FTPI-MINN-10/20 \\
UMN-TH-2912/10 \\
August 2010 \\
\end{flushright}
\vspace{0.3in}
\begin{center}
{\Large \bf Neutrino scattering on atomic electrons in searches for neutrino magnetic moment 
\\}
\vspace{0.2in}
{\bf M.B. Voloshin  \\ }
William I. Fine Theoretical Physics Institute, University of
Minnesota,\\ Minneapolis, MN 55455, USA \\
and \\
Institute of Theoretical and Experimental Physics, Moscow, 117218, Russia
\\[0.2in]
\end{center}

\begin{abstract}

The scattering of a neutrino on atomic electrons is considered in the situation where the energy transferred to the electrons is comparable to the characteristic atomic energies, as relevant to the current experimental search for neutrino magnetic moment. The process is contributed by the standard electroweak interaction as well as by the possible neutrino magnetic moment. Quantum mechanical sum rules are derived for the inclusive cross section at a fixed energy deposited in the atomic system, and it is shown that the differential over the energy transfer cross section is given, modulo very small corrections, by the same expression as for free electrons, once all possible final states of the electronic system are taken into account. Thus the atomic effects effectively cancel in the inclusive process.
\end{abstract}

\end{titlepage}

The magnetic moments of neutrinos in the Standard Model are proportional to neutrino masses~\cite{fs} and are very small: $\mu_\nu \, \ltap \, 10^{-20} \, \mu_B$ with $\mu_B = e/(2m_e)$ being the Bohr magneton. Thus any evidence of a significantly larger neutrino magnetic moment
(NMM) would undoubtedly reveal effects of new physics. 

The current experimental limits for reactor (anti)neutrinos are provided by the dedicated experiments TEXONO~\cite{tx} and GEMMA~\cite{ge1,ge2} with the latest upper limit~\cite{ge2} being $\mu_\nu < 3.2 \times 10^{-11} \, \mu_B$. Both experiments measure the energy $T$ deposited in ultra low background Germanium crystal detectors exposed to neutrino flux from a reactor. In a scattering of neutrino with energy $E_\nu$ off a free electron the energy $T$ is the kinetic energy of the recoiling electron, and the differential over $T$ cross section is given by the incoherent sum of the scattering due to the NMM
\be
{d \sigma_{(\mu)} \over dT }= 4 \pi \, \alpha \,  \mu_\nu^2 \, \left ( {1 \over T} - {1 \over E_\nu } \right ) = \pi \, {\alpha^2 \over m_e^2}  \, \left ( {\mu_\nu \over \mu_B} \right )^2 \, \left ( {1 \over T} - {1 \over E_\nu } \right )
\label{fe}
\ee
and a constant in $T$ (at $T \ll E_\nu$) contribution from the standard electroweak interaction
\be
{d \sigma_{EW} \over dT }= {G_F^2 \, m_e \over 2 \pi} \left ( 1+ 4 \, \sin^2 \theta_W + 8 \, \sin^4 \theta_W \right ) \, \left [ 1 + O \left ( {T \over E_\nu} \right) \right ] \approx 5 \times 10^{-48} \, {\rm cm^2 / keV}.
\label{sew}
\ee
(A compilation and discussion can be found e.g. in Ref.~\cite{ve}.) 

Due to the $1/T$ singularity of the electromagnetic scattering, an improvement of the upper bound on NMM requires going down to a lower threshold in the energy deposited in the detector, and the most recent experiments have this threshold in the range of a few keV. Such energies however are comparable to the characteristic atomic energies $\varepsilon_0$ in $Ge$, for which a representative value can be that of the $K_\alpha$ line 9.89\,keV. Clearly, in this situation it is legitimate to question applicability of the formulas (\ref{fe}) and (\ref{sew}) derived for free electrons, and atomic effects should be taken into account. In particular it has been argued~\cite{wll} that the atomic effects in $Ge$ very significantly enhance the NMM scattering cross section in the keV energy range. 

The present paper revisits the issue of the atomic effects on the neutrino scattering. It will be shown that in the relevant range of low excitation energy $T$ the cross section summed over the final states of the electrons is governed by  quantum-mechanical sum rules and the inclusive cross section per atomic electron is essentially given by {\em unmodified} formulas in the equations (\ref{fe}) and (\ref{sew}). 

We start with considering in detail the more interesting case of the electromagnetic scattering due to NMM, and then extend the treatment to the standard electroweak process. Let $k_\mu$ and $k'_\mu$ be the four-momenta of the initial and the final neutrino, so that $q=k-k'$ is the four-momentum transferred to the atomic system, $q=(T, \vec q)$. It is assumed that $T$ is much less than the energy $E_\nu$ of the incoming neutrino, so that $E'_\nu \approx E_\nu$, and also that $T$ is much smaller than the electron mass, $T \ll m_e$, so that the electrons in the atom and in the scattering can be treated within nonrelativistic quantum mechanics~\footnote{This is quite similar to the treatment of the inelastic scattering of fast electrons on atoms as can be found e.g. in the textbook~\cite{ll}}. One can also notice that in the energy range of interest for current experiments the condition $T \gg \varepsilon_0 \, m_e/M$ is satisfied, which allows to assume that no energy is spent on the recoil of the atom as a whole including its nucleus with the mass $M$.  The nucleus is thus considered to be infinitely massive and at rest, so that the interaction with it makes no contribution to the scattering at the energy transfer $T$, and only the interaction with the atomic electrons is of relevance. It is also implied that $T$ is above the ionization threshold, so that it is the processes with emission of electron(s) in the continuum that contribute to the cross section, rather than just an excitation of discreet atomic levels. 

The NMM interaction with the electromagnetic field $A_\mu(q)$ of the electrons is described by the standard term in the Lagrangian
\be
L_{NMM}= \mu_\nu \left ({\overline \nu}(k') \, \sigma_{\mu \nu} \, \nu(k) \right ) \, q_\mu \, A_\nu~.
\label{lmm}
\ee
In the leading nonrelativistic order the electrons only create a Coulomb field, whose potential $A_0$ is given by $A_0(\vec q)=\sqrt{4 \pi \alpha} \, \rho( \vec q)/ {\vec q}^{\,2}$ with $\rho(\vec q)$ being the Fourier transform of the electron number density operator
\be
\rho(\vec q)= \sum_{a=1}^Z \exp(i \vec q \cdot \vec r_a)~,
\label{ne}
\ee
and the summation runs over the positions $\vec r_a$ of all the $Z$ electrons in the atom. It is a straightforward exercise to find the cross section for scattering on the ground state of the atom due to the interaction (\ref{lmm}) in the form
\be
{d^2 \sigma_{(\mu)} \over dT \, d Q^2} = 4 \pi \, \alpha \, { \mu_\nu^2 \over Q^2} \, \sum_n \, \delta (T - E_n+E_0) \, \left | \langle n | \rho(\vec q) | 0 \rangle \right |^2~,
\label{d2s}
\ee
where $Q^2=\vec q^{\, 2}$,  the sum runs over all the states $| n \rangle$ of the electron system with $|0 \rangle$ being the ground state, and $E_n$ stands for the energy of the corresponding state. One can readily reproduce the $1/T$ term\footnote{Clearly the $1/E_\nu$ term in Eq.(\ref{fe}) is neglected in the considered approximation.} in Eq.(\ref{fe}) for the scattering on free electron by noticing that in this case the sum in Eq.(\ref{d2s}) contains only one term (corresponding to a free electron with momentum $\vec q$)   and is equal to $ \delta (T - Q^2/2m_e)$, so that the integration over $Q^2$ is trivial.

One can further notice that the sum in Eq.(\ref{d2s}) is proportional to $Q^2$ at low momentum transfer, i.e. when $|\vec q|$ is smaller than the characteristic momenta of the electrons in the atom, $Q^2 \ll 2 \, m_e \, \varepsilon _0$. In this limit the exponent in the expression (\ref{ne}) can be expanded in the Taylor series, and the unit term gives no contribution due to the orthogonality of the ground and excited states. Keeping the first nonvanishing term one finds
\be
{d^2 \sigma_{(\mu)} \over dT \, d Q^2} = 4 \pi \, \alpha \,  \mu_\nu^2  \, \sum_n \, \delta (T - E_n+E_0) \, \left | \langle n |d_x | 0 \rangle \right |^2~,
\label{dds}
\ee
where $d_x$ is the projection on the direction of $\vec q$ of the dipole operator $\vec d = \sum_a \vec r_a$. The formula in Eq.(\ref{dds}) can be used at $Q^2 = T^2$, i.e. for the on-shell photon, to relate the discussed cross section to that of the photoelectric effect for a real photon with energy $T$: $\sigma_\gamma(T)$. The latter cross section is determined by the same sum over the dipole matrix elements (see e.g. in the textbook~\cite{blp}), so that one finds~\cite{wll}
\be
\left . {d^2 \sigma_{(\mu)} \over dT \, d Q^2}\right |_{Q^2=T^2} = {\mu_\nu^2 \over \pi} \, {\sigma_\gamma(T) \over T}~.
\label{nuga}
\ee
This relation is however of little help in finding a reliable approximation for the integral over $Q^2$ that is necessary for determining the experimentally measured full inclusive cross section $d \sigma_{(\mu)}/dT$. The reason is that the integral receives contribution from the regions of $Q^2$ where the photon momentum is comparable to  the characteristic atomic momenta as well as from the overlapping at $T \sim \varepsilon_0$ region where $Q^2 \approx 2m_e \, T$. At those $Q^2$ the dipole approximation is no longer valid~\footnote{An integration in Eq.(\ref{nuga}) over all kinematically allowed values of $Q^2$ i.e. up to $Q^2 \approx 4 \,E_\nu^2$ without introducing a form factor leads to the claim~\cite{wll} of a giant enhancement of the cross section by atomic effects, but is clearly unjustified since the sum in Eq.(\ref{d2s}) rapidly falls off at large $Q^2$. The relation (\ref{nuga}) however can be of use in situations where $E_\nu$ is small in the scale of the characterisc size of the target system, such as in the problem of deuteron splitting by reactor or solar neutrinos~\cite{ab}. }.  

The full integral of the expression in Eq.(\ref{d2s}) over $Q^2$ can be found using a quantum-mechanical sum rule. Indeed, the sum in that expression can be written in terms of the imaginary part the function $R(T,Q^2)$:
\be
\sum_n \, \delta (T - E_n+E_0) \, \left | \langle n | \rho(\vec q) | 0 \rangle \right |^2={1 \over \pi} {\rm Im} R(T,Q^2)~,
\label{suim}
\ee
with
\be
R(T,Q^2)=\sum_n {1 \over T - E_n+E_0 - i \, \epsilon} \, \left | \langle n | \rho(\vec q) | 0 \rangle \right |^2 = \left \langle 0 \left |\rho(- \vec q) \, {1 \over T-H+E_0- i \, \epsilon}\, \rho(\vec q) \right | 0 \right \rangle~,
\label{rdef}
\ee
where $i \epsilon$ is, as usually, an infinitesimal shift from the the real axis, and $H$ is the full Hamiltonian for the atomic electrons. At a fixed $T$ and generally complex $Q^2$ the function $R(T,Q^2)$ is an analytic function of $Q^2$ with a cut along the positive real axis, and this function is manifestly real at real negative $Q^2$, so that $R(T,z^*)=R^*(T,z)$, and its imaginary part on the cut vanishes at $Q^2 \to 0$, as is explained in the above discussion leading to Eq.(\ref{dds}). At large $Q^2$ this function is determined by the final states of electrons with large momenta, where the atomic effects are negligible, so that it falls at large $|Q^2|$ as
\be
R(T,Q^2) \to -Z \, {2 \, m_e \over Q^2}~,~~~~(|Q^2| \to \infty)~.
\label{rinf}
\ee
One thus concludes that the function $R$ satisfies the dispersion relation with no subtractions
\be
R(T,P^2)={1 \over \pi} \, \int_0^{\infty} \, {{\rm Im} R(T, Q^2) \over Q^2 - P^2 - i \epsilon} \, d Q^2~.
\label{disp}
\ee
Consider now the limit $P^2 \to 0$. The operator $\rho(\vec p)$ at $p \to 0$ becomes a unit operator for each electron, so that only the ground state contributes to the sum in Eq.(\ref{rdef}) and one finds $R(T,0) = Z/T$. Upon substituting $P^2 \to 0$, the dispersion relation (\ref{disp}) thus yields
\be
{1 \over \pi} \, \int_0^{\infty} \, {\rm Im} R(T, Q^2) \, {dQ^2 \over Q^2} = {Z \over T}~.
\label{rint}
\ee
Given the relation (\ref{suim}), this integral is almost exactly what one needs to calculate the inclusive differential cross section $d \sigma_{(\mu)}/dT$, except that in the latter calculation the integral runs within the kinematical limits for $Q^2$, i.e. from $Q^2=T^2$ to $Q^2 \approx 4 E_\nu^2$, rather than from zero to infinity. By our assumptions the neutrino energy is much larger than either the atomic scale or $T$, so that within our approximation the scale $E_\nu^2$ in the upper limit is indistinguishable from infinity. As to the lower limit, the difference between the two integrals is obviously given by the integral from zero to $Q^2=T^2$. In this range (and at $T \sim \varepsilon_0$) one can safely use the dipole approximation described by Eq.(\ref{dds}) and thus conclude that the integral describing the difference in the lower integration limits can be estimated in terms of the characteristic atomic size $r_0$ as being of the order of $(T^2 \, r_0^2) \,  Z/T$ and is much smaller than $Z/T$. Thus, up to this parametrically small difference, the inclusive cross section is determined by the sum rule (\ref{rint}) and is given (per electron) by the $1/T$ term in the expression (\ref{fe}) derived for a free electron\footnote{It can be also noticed that both small corrections due to the kinematical restrictions are negative, so that the full integral in Eq.(\ref{rint}) in fact provides an upper bound on $d \sigma_{(\mu)}/dT$}. 

Proceeding to discussion of the standard electroweak scattering, it can be noted that similarly to Eq.(\ref{d2s}) the double differential cross section can be readily expressed in terms of the imaginary part of the function $R(T,Q^2)$ as 
\be
{d^2 \sigma_{EW} \over dT \, d Q^2} = {G_F^2 \over 4 \pi} \left ( 1+ 4 \, \sin^2 \theta_W + 8 \, \sin^4 \theta_W \right ) \, \left [ {1 \over \pi} \, {\rm Im} R(T,Q^2) \right ]~,
\label{d2sw}
\ee
so that the function ${\rm Im} R(T,Q^2)$ enters with a constant, rather than $Q^{-2}$, weight, as it should be for a point-like interaction.
The sum rule for the full integral of this function over $Q^2$ immediately follows from considering the dispersion relation (\ref{disp}) at $P^2 \to -\infty$ and  comparing it with the asymptotic expression in Eq.(\ref{rinf}). In this way one finds
\be
{1 \over \pi} \, \int_0^\infty {\rm Im} R(T,Q^2) \, dQ^2 = 2 \, Z \,  \, m_e~.
\label{sr2}
\ee
In this case the contribution of the `extra' integration region of $Q^2 < T^2$ near the lower limit is of the relative order $(T/m_e) \, (T \, r_0)^2$ and is much less than the relativistic corrections so that it can safely be neglected. One thus can readily perform the integration of the expression in Eq.(\ref{d2sw}) over the kinematical range of $Q^2$ and arrive at the same formula for the inclusive differential cross section per electron $d\sigma_{EW} / dT$ as given by the the free electron relation (\ref{sew}).

The existence of the simple expressions for the integrals in the sum rules (\ref{rint}) and (\ref{sr2}) is quite specific to the weight functions $Q^{-2}$ and ${Q^0}$ in those integrals, and generally one would not expect similarly simple relations for other weight factors. (One such `other' weight function $Q^{-4}$ appears in the well known case of ionization by fast charged particles~\cite{ll}.) The origin of the sum rules (\ref{rint}) and (\ref{sr2}) with the respective weight functions can be somewhat clarified by considering a simple example of scattering on the ground state of one electron moving in a spherically symmetric potential $V(r)$. The Hamiltonian for the electron thus has the form
$H(\vec p,\vec r)=\vec p^{\,2}/ 2 m_e + V(r)$,
and  the function $R(T,Q^2)$ can be written as
\begin{eqnarray}
R(T,Q^2) &=& \left \langle 0 \left |\, e^{-i \vec q \cdot \vec r} \, \left [ T -H(\vec p, \vec r) + E_0 \right ]^{-1} \,  e^{i \vec q  \cdot \vec r}\, \right | 0 \right \rangle = \nonumber \\
\left \langle 0 \left | \, \left [ T -H(\vec p + \vec q, \vec r) + E_0  \right ]^{-1} \,  \right | 0 \right \rangle &=&
 \left \langle 0 \left | \, \left [ T -{\vec q^{\,2} \over 2 m_e}- {\vec p \cdot \vec q \over m_e}  - H(\vec p, \vec r) + E_0  \right ]^{-1} \,  \right | 0 \right \rangle \, ,
\label{r1}
\end{eqnarray}
where the infinitesimal shift $T \to T - i \epsilon$ is suppressed for brevity. 

Consider now a formal expansion of the latter expression in inverse powers of $(T-Q^2/2 m_e)$ and consider first the resulting terms containing powers of $(\vec p \cdot \vec q)$. Clearly, the terms with odd powers of this scalar product vanish upon averaging due to parity. The terms with even powers after averaging over the state $| 0 \rangle$ result in expressions of the generic form 
\be
{(Q^2)^u \over \left ( T - {Q^2 \over 2 m_e} \right )^w }\, \Phi_{u, w}
\label{geneven}
\ee
with $\Phi_{u, w}$ being the coefficients arising from the averaging of the operators depending on even powers of $\vec p$ and $\vec r$ in the corresponding terms of the expansion. It is important that the integer powers $u$ and $w$ in the expression (\ref{geneven}) satisfy the inequality
\be
w \ge 2 \, u +1~.
\label{wuineq}
\ee
Using the formula 
\be
{1 \over \pi} \int \, {\rm Im} \left [{ (Q^2)^s \over \left ( T - {Q^2 \over 2 m_e} - i \epsilon \right )^w } \right ] \, d Q^2 = {2 m_e \over (w-1)!} \left .  \left ( 2 m_e  {d \over dQ^2} \right )^{w-1}\, (Q^2)^s  \right |_{Q^2 = 2 m_e T}~,
\label{im0}
\ee 
which gives zero at $w \ge s+2$ and both $s$ and $w$ being integer, one can readily see that, due to the condition (\ref{wuineq}), all the nontrivial terms of the expansion in $(\vec p \cdot \vec q)$ give no contribution to the integrals in the l.h.s. of the equations (\ref{rint}) and (\ref{sr2}). We thus conclude that for the purpose of calculating the integrals in Eqs.(\ref{rint}) and (\ref{sr2}) the expression in Eq.(\ref{r1}) can be replaced by a much simpler one, where the product $(\vec p \cdot \vec q)$ is omitted:
\be
\left \langle 0 \left | \, \left [ T -{\vec q^{\,2} \over 2 m_e}  - H(\vec p,\vec r) + E_0 -i \epsilon  \right ]^{-1} \,  \right | 0 \right \rangle =  \left ( T -{Q^{2} \over 2 m_e} -i \epsilon \right )^{-1}~,
\label{simpf}
\ee
which immediately results in the integrals satisfying the sum rules (\ref{rint}) and (\ref{sr2}).

The one electron example illustrates the reason for the importance of the specific weight factors in the considered here integrals relevant to the neutrino scattering: for weight functions with larger {\it positive} powers of $Q^2$ some terms of the expansion in 
$(\vec p \cdot \vec q)$ give a nonzero contribution, while for higher {\it negative} powers of $Q^2$, as in the ionization by charged particles, the integrals are generally divergent at the lower limit~\footnote{The latter divergence is dominated by the $Q^2$ behavior of ${\rm Im} R(T,Q^2)$ at low $Q^2$, which, as discussed, can be expressed in terms of the photoelectric cross section.}.

As a general remark, it can be noticed that the discussed here treatment of the inclusive scattering on the atom is in a close analogy with the well developed approach to the deep inelastic scattering (DIS). However the nonrelativistic dynamics of the target brings a great simplification, and it is easier to directly derive the necessary sum rules using quantum mechanics, rather than by fully using the analogy with similar relations in DIS. As discussed here, the derived in this way sum rules (\ref{rint}) and (\ref{sr2}) determine that both for the hypothetical NMM interaction and for the standard electroweak one the inclusive differential in $T$ cross section per electron is essentially not affected by the atomic effects down to quite low values of the energy transfer $T$, well within the range of interest for the current neutrino experiments. 

I thank A.S. Starostin for alerting me to the relevance of the discussed problem to the current experimental searches, and I acknowledge a helpful discussion with M. Shifman. This paper was finalized at the Aspen Center for Physics. This work is supported in part by the DOE grant DE-FG02-94ER40823.

\end{document}